\documentclass[aps,prb,superscriptaddress,twocolumn]{revtex4-2}
\pdfoutput=1
\usepackage{graphicx}
\usepackage{dcolumn}
\usepackage{bm}
\usepackage{breqn}
\usepackage{physics}
\usepackage{hyperref}
\usepackage{mathptmx}
\usepackage[english]{babel}
\usepackage{amssymb}
\usepackage{amsfonts}
\usepackage{amsmath}
\usepackage{eucal}
\usepackage{graphicx}
\usepackage{epsfig}
\usepackage{appendix}
\usepackage[dvipsnames]{xcolor} 
\usepackage{bbm}
\usepackage{subfigure,dcolumn}
\usepackage[T2A,T1]{fontenc}
\usepackage{braket}
\usepackage[colorinlistoftodos]{todonotes}
\usepackage[utf8]{inputenc}
\usepackage{amsthm}
\usepackage{amsmath}
\usepackage{slashed}
\usepackage[normalem]{ulem}
\hypersetup{colorlinks=true}

\begin{document}
\newcommand{\commentD}[1]{{\color{Purple} {\bf DZ:} #1}}
\newcommand{\commentJ}[1]{{\color{Blue} {\bf JM:} #1}}



\newcommand{\be}{\begin{equation}}
\newcommand{\ee}{\end{equation}}
\newcommand{\bea}{\begin{eqnarray}}
\newcommand{\eea}{\end{eqnarray}}
\newcommand{\ba}{\begin{align}}
\newcommand{\ea}{\end{align}}

\def\la{\langle}
\def\ra{\rangle}

\def\cB{\mathcal{B}}
\def\cD{\mathcal{D}}
\def\cE{\mathcal{E}}
\def\cU{\mathcal{U}}
\def\cN{\mathcal{N}}
\def\cK{\mathcal{K}}
\def\cR{\mathcal{R}}
\def\cO{\mathcal{O}}
\def\cH{\mathcal{H}}
\def\cF{\mathcal{F}}
\def\cS{\mathcal{S}}

\def\va{{\bf a}}
\def\vb{{\bf b}}
\def\vc{{\bf c}}
\def\vd{{\bf d}}
\def\ve{{\bf e}}
\def\vf{{\bf f}}
\def\vg{{\bf g}}
\def\vh{{\bf h}}
\def\vr{{\mathbf{r}}}
\def\vx{{\mathbf{x}}}
\def\vF{{\mathbf{F}}}
\def\vB{{\mathbf{B}}}
\def\vp{{\mathbf{p}}}
\def\vq{{\mathbf{q}}}
\def\vk{{\mathbf{k}}}
\def\vs{{\mathbf{s}}}


\title{Non-Gaussian Variational Wavefunctions for Interacting Bosons on the Lattice}
\author{T. Qian}
\affiliation{
College of Physics and Optoelectronic Engineering. Ocean University of China.}
\author{J. J. Fern\'andez-Melgarejo}
\affiliation{
Departamento de  Electromagnetismo y Electr\'onica. Universidad de Murcia.}
\author{D. Zueco}
\affiliation{
Instituto de Nanociencia y Materiales de Arag\'on (INMA)- CSIC. Universidad de Zaragoza.}
\author{J. Molina-Vilaplana}
\affiliation{
Departamento de Autom\'atica. Universidad Polit\'ecnica de Cartagena.}


\begin{abstract}
A variational method for studying the ground state of strongly interacting quantum many-body bosonic systems is presented. Our approach constructs a class of extensive variational non-Gaussian  wavefunctions which extend Gaussian states by means of nonlinear canonical transformations (NLCT)  on the fields of the theory under consideration.  We illustrate this method with the one dimensional Bose-Hubbard model for which the proposal presented here, provides a family of  approximate ground states at arbitrarily large values of the interaction strength. We find that, for different values of the interaction, the non-Gaussian NLCT-trial states sensibly improve the ground state energy estimation when the system is in the Mott phase. 
\end{abstract}

\maketitle

\section{Introduction}
\label{sec:intro}
One of the major problems in many body physics is to understand the phenomena associated to strongly coupled systems. This includes a huge variety of effects ranging from quark confinement  to strongly correlated electron systems in condensed matter physics. To do so, nonperturbative methods are required. Many useful techniques such as path integral approaches, large $N$ expansions or numerical methods (such as density matrix renormalization group or tensor networks) have been developed to address these problems. 
While these methods have successfully been applied to a broad range of problems,  the complexity of the theoretical proposals have obscured the understanding of the physical phenomena. In this sense, it is acknowledged that the use of  variational methods allow to tackle these problems to some extent in a relatively simple way by means of variational wavefunctions. When it comes to nonperturbative aspects, there are situations in which the use of wavefunctionals  exhibits clear advantages. For example, path integral methods are especially suited to compute quantities that have no perturbative contributions and can be addressed through a saddle point approximation. Nevertheless, in case the observables of interest can receive both perturbative and nonperturbative contributions, the path integral approach becomes more difficult \cite{Moshe:2003xn}.

Choosing appropriate variational wavefunctions for a strongly interacting many body system is a challenging issue \cite{Kogan:1994wf}. Firstly, one has the problem of the \emph{generality} of the trial state. Namely, the trial state should be general enough to capture the most salient physical features of the phenomena. Due to the enormous size of the Hilbert space in a many body system, it is very difficult to identify by mere intuition the relevant features that have to be grasped by the \emph{ansatz}. Thus, a systematic method to build them would be desirable. Secondly, one must face the problem of \emph{calculability}. Even  possessing a reasonable and flexible \emph{ansatz} for the  wavefunctional, one wants to evaluate expectation values of operators/observables of interest in this state which, in general, will be challenging.
Given the very limited ability to evaluate expectation values with non-Gaussian wavefunctionals, the calculability requirement on the trial wave functional has constrained the form of the trial wavefunctionals to Gaussian states. 

Gaussian states  are given by the exponentials of quadratic functionals of creation and annihilation operators of the fundamental fields of the theory under consideration. The expectation values of physical observables can be efficiently computed for these states \cite{Zinn-Justin:1996khx}, which obey Wick's theorem, thus allowing to express expectation values of arbitrary products of mode operators in terms of products of pairs \cite{Wick50}.  Gaussian approximations as the  Hartree–Fock–Bogoliubov one are  used to approximate the dynamics of interacting bosons \cite{Quijandra2015, Naether2015}.
While Gaussian states represent the exact ground state in noninteracting systems, they have some important limitations in capturing phenomena related to interacting systems.
Therefore, it is interesting to extend those ansatze to non-Gaussian generalizations. Given, however, the very limited ability to evaluate expectation values with non-Gaussian wavefunctionals, the calculability requirement on the trial wave functional is typically restricted to the form of the trial wavefunctionals to Gaussian states. 

Finally, one must address the problem of the  ultraviolet modes. The main objective of a variational calculation in a strongly interacting  system is to obtain the correct configuration for the low momentum modes of the field in the vacuum wavefunctional. Due to the interaction between the high and low momentum modes in an interacting system it is thus desirable  to have a method that yields  variational parameters that optimally integrate out the effects of high energy modes into the low energy physics.

In this work, following the pioneering works \cite{Polley:1989wf, Ritschel:1990zs,IbanezMeier:1991hm} we develop a class of variational non-Gaussian  wavefunctions which extend Gaussian states by means of nonlinear canonical transformations (NLCT) on the fields of the theory under consideration.  The presented scheme is self-consistent and requires no other assumptions than the choice of a NLCT-variational manifold. These variational wavefunctions, has been shown to possess strong entanglement between the microscopic degrees of freedom \cite{Fernandez-Melgarejo:2020utg}, while retaining most of the calculability of Gaussian wavefunctions. There exist several related methods, like the cluster expansion \cite{Hsue:1985ke,Schutte:1985sd}, the $t$-expansion \cite{Horn:1984bq}, and the
$\delta$-expansion \cite{Bender:1988rq}, which, even going (slightly) beyond the Gaussian ansatz, they yield expectation values that cannot be exactly calculated and have to be approximated by additional series expansions.

We illustrate the NLCT method with the Bose-Hubbard model in one dimension. Our approach provides a family of  approximate ground states at arbitrarily large values of the interaction strength. Quantum interacting bosons in 1D provide an exciting area where condensed matter, low-temperature, and ultracold atomic converge.  The Bose-Hubbard (BH) model describes a system of interacting spinless bosons on the lattice. This model provides a theoretical description of interacting cold atoms in optical lattices \cite{Cazalilla_2011,Krutitsky_2016}. Besides, some numerical solutions  exist for the BH model, that allows us to benchmark our non-Gaussian ansatze \cite{Cazalilla_2011}.

In line with our approach, in  Ref. \cite{Shi_2018}, authors built non-Gaussian wavefunctions for many body systems in the lattice using a set of unitary transformations on Gaussian states. The nature of the transformation depends on the theory under consideration. Its general form was inspired by canonical transformations in condensed matter physics, such as the polaron transformations in electron-phonon systems and flux attachment in FQHE. While the construction in \cite{Shi_2018} may work for mixed boson-fermion systems, it offers a very limited class on purely non-Gaussian bosonic states. This is due to the challenge posed by what we call the \emph{truncation} problem for bosons that will be discussed later in this paper.

In the context of high energy physics, NLCT-wavefunctions have been recently used to build a nonperturbative version of continuous entanglement renormalization tensor networks in order to explore connections between tensor networks and the AdS/CFT holographic correspondence \cite{Fernandez-Melgarejo:2019sjo, Fernandez-Melgarejo:2020fzw,Fernandez-Melgarejo:2021mza}. The tensor network circuit there, implements a series of scale-dependent NLCT. It was shown that the leading contribution to the entanglement entropy, comes from the Gaussian part of the ansatz and are always related to the leading area term in the holographic calculation. On the other hand, the subleading contributions are given by the non-Gaussian part of the ansatz and are related with quantum corrections to the holographic entanglement entropy.

The paper is structured as follows:  A description of the NLCT transformation method and some technical aspects, as the calculation of expectation values, are presented in Section \ref{sec:nggva}. In Section \ref{sec:gva} we introduce the 1D Bose-Hubbard model and briefly review the variational approach to its ground state using Gaussian states (Gaussian Variational Approach, GVA). In Section \ref{sec:nlct_bh} we detail the calculation of the energy expectation value for the BH Hamiltonian in a concrete NLCT-Non-Gaussian state.  We discuss on the suitability of this choice of NLCT-state for the problem at hand and provide expressions for the energy functional that will be subsequently optimized. Section \ref{sec:results} details the optimization of the ground state energy functional and the results where, our approximation to the ground state of the 1D Bose-Hubbard model for arbitrarily large values of the interaction strength, shows a sensible improvement on the estimation of the ground state energy with respect to the Gaussian case, especially when the system is in the Mott phase. We finish with a discussion on the results and an outlook in Section \ref{sec:discussion}.

\section{Non Gaussian Variational Ansatz}\label{sec:nggva}

In a  bosonic QFT on a lattice, let us consider a normalized Gaussian variational wavefunctional
\begin{align}
    \Psi_G(\lambda)\equiv \mathcal{N}\, \exp \left(-\frac{1}{2} \sum_{k,l}\, b_k^\dagger\, \Lambda_{k\, l} b_l\right)\, ,
    \label{eq:generalized_cs}
\end{align}
where the creation and annihilation operators for the bosonic modes are denoted by  $b^{\dagger}_k$ and $b_k$ respectively such that $[b_k,b^{\dagger}_{k'}]=\delta_{k,k'}$, and $\Lambda_{k l}$ is a matrix depending on the variational parameters $\lambda$. Then, extensive non-Gaussian trial  wavefunctions can be non-perturbatively built as
\be\label{eq:NG_wavefunc}
|\Psi(\lambda,h)\ra_{NG} = \cU(h)\, |\Psi(\lambda)\ra_{G}=\exp(\cB(h))\, |\Psi(\lambda)\ra_{G} \, ,
\ee
with $\cU(h)=\exp(\cB(h))$, and $\cB(h)^{\dagger} = -\cB(h)$ an anti-Hermitian non-quadratic operator (in terms of $b^{\dagger}_k$ and $b_k$ ) that non-perturbatively adds new variational parameters labelled by $h$ to those $\lambda$ defining $|\Psi(\lambda)\ra_{G}$.

 The exponential nature of $\cU$ ensures the correct extensive volume dependence of observables and specifically the total energy of the system. As $\cU$ is unitary, the normalization of the state is not affected. It is straightforward to see that the expectation value of any operator $\cO$ with $|\Psi(\lambda,h)\ra_{NG}\equiv |\Psi\ra_{NG}$ amounts to the calculation of a Gaussian expectation value for the transformed operator $\tilde{\cO}=\cU^{\dagger}\, \cO\, \cU$,
 \be\label{eq:nlct_operator}
 \la \Psi|{\cO}|\Psi\ra_{NG}\equiv \la \Psi|\tilde{\cO}|\Psi\ra_{G}\, .
 \ee

In principle, for bosonic systems, both in finite and infinite lattices, any non-quadratic choice for $\cB(h)$, while  leading  to a non-Gaussian trial state, it induces an infinite commutator expansion via the Hadamard's lemma
\begin{align}
\tilde{\cO} = e^{{\rm ad}_{\cB}}\, \cO\, 
&= \sum_{n=0}^{\infty}\, \frac{(-1)^n}{n!}\, \left[\cB, \cO \right]_n\, , \\
\left[\cB, \cO \right]_n &\equiv \underbrace{\left[\cB\, \left[\cB,\cdots \cO\right]\right]}_{n\,  {\rm times}} \, .
\label{eq:conm_series}
\end{align}
This spoils any possibility on having an ansatz with finite \emph{calculability} properties as any computation of an expectation value, amounts to the evaluation of an infinite series of Gaussian expectation values. This is what we call the \emph{truncation} problem. In fermionic systems in finite lattices, this basic problem is alleviated by the anticonmuting nature of its operator algebra. Namely, for a non-quadratic fermionic operator $\mathcal{F}(h)$, 
\be
\exp\left[ \mathcal{F}(h)\right]\, ,
\ee
has only a finite number of terms in its formal power expansion series or, equivalently, $\left[\mathcal{F}, \cO \right]_n = 0$ at least for $n \geq N$, with $N$ the number of sites in the lattice. This of course does not ensure that any further structure is needed in order to find fermionic ansatze with nice calculability properties (see \cite{Fernandez-Melgarejo:2020fzw} for examples in fermionic field theory), but it is worth to mention that the truncation problem is especially ill-posed in bosonic systems. 


Remarkably, the method presented here, provides an ansatz for bosonic systems that automatically implements a controllable truncation in Eq.(\ref{eq:conm_series}). This reduces the calculation of expectation values of operators to a finite number of Gaussian expectation values. The operator $\cB$ consists of a product of bosonic operators $\pi$'s and $\phi$'s, which is given by
\be\label{eq:BBos}
\cB
= 
-s\int_{p,\lbrace q_i\rbrace}
h_{p q_1\cdots q_m} \, \pi_p\, \phi_{q_1}\ldots\phi_{q_m} \,  \delta_{p,-\sum q_i} 
\ ,
\ee
with $\phi_k=\frac{1}{\sqrt{2}}\left ( b_k  + b^{\dagger}_{-k} \right )$, $\pi_k=\frac{1}{\sqrt{2}\, i}\left ( b_{-k}  -  b^{\dagger}_{-k} \right )$ in such a way that $ [\phi_k, \pi_{k^\prime} ] = i \delta_{k, -k^\prime}$ and $m \in \mathbb{N}$.
We will denote these operators from here in advance symbolically  as $\cB\equiv \pi\, \phi^{m}$. Here, $s$ is a  variational parameter that tracks the deviation of any observable from the Gaussian case. $h_{p,q_1,\ldots,q_m}$ is a variational function that must be optimized upon energy minimization. It is symmetric w.r.t. exchange of $q_i$'s  and is constrained to satisfy:

\begin{eqnarray}
h_{p,q_1,\ldots,q_m}
=0 
\ , 
\quad p=q_i\, ,\\ \nonumber
h_{p,q_1,\ldots,q_m}\, \times h_{q_i,k_1,\ldots,k_m} = 0
\, .
\label{eq:constraint}
\end{eqnarray}

These conditions ensure that the commutator series (\ref{eq:conm_series}) terminates after the first nontrivial term. Namely, the constraints in  (\ref{eq:constraint}) are the responsible for this truncation when the Hadamard's lemma is applied.
The action of $\cU$ on the canonical field operators $\phi_k$ and $\pi_k$ is given by
\bea
\label{trasfields}
	\tilde \phi_k&\equiv&\cU^\dagger\, \phi_k\, \cU 
	=
	\phi_k
	+s \, \Phi_k
	\ ,
	\\ \nonumber
	\tilde \pi_k&\equiv&\cU^\dagger\, \pi_k\, \cU
	=
\pi_k
	-\, s \, \Pi_k
	\ ,
\eea
where $\Phi_k,\, \Pi_k$ are defined as the nonlinear field functions, 
\begin{equation} \label{detail_transfields}
\begin{aligned}
\Phi_k\equiv&\ \int_{q_i} h_{k,q_1,\ldots,q_m}\, \phi_{q_1}\cdots \phi_{q_m} \delta_{p, \sum q_i} \ , 
\\ 
\Pi_k
	\equiv&\
	m\, \int_{q_i}
	h_{q_1,k,\ldots,q_m}\, \pi_{q_1}\, \phi_{q_2}\phi_{q_m}
	\delta_{p, \sum q_i}
\ .
\end{aligned}
\end{equation}
Being $\cU$ unitary, the canonical commutation relations (CCR) still hold under the nonlinear transformation of the fields, giving  $[\tilde \phi_p, \tilde \pi_q]=i\delta_{p,-q}$.  For this reason, the above transformations are known as nonlinear canonical transformations (NLCT).
\medskip

Regarding observables, it is of particular interest to consider $n$-point correlation functions $\expval{\phi_{k_1}\cdots\phi_{k_n}}_{NG}$. To evaluate this, we use \eqref{trasfields} and \eqref{detail_transfields}, to obtain

\begin{widetext}
\begin{eqnarray}\label{eq:corr_funcs}
\expval{\phi_{k_1}\cdots\phi_{k_n}}_{NG} =\expval{\phi_{k_1}\cdots \phi_{k_n}}
	+ s\left [\expval{\Phi_{k_1}\phi_{k_2}\cdots \phi_{k_n}} +\cdots +\expval{\phi_{k_1}\cdots \phi_{k_{n-1}}\Phi_{k_n}}\right]
	\\ \nonumber
	+s^2\left[ \expval{\Phi_{k_1}\Phi_{k_2}\phi_{k_3}\cdots \phi_{k_n}}
	+\cdots +\expval{\phi_{k_1}\cdots\Phi_{k_{n-1}}\Phi_{k_n}}\right]
\cdots
+ s^n \expval{\Phi_{k_1}\cdots\Phi_{k_n}}\, ,
\end{eqnarray}
\end{widetext}
where $\la \cdots \ra$ refers to a an expectation value taken w.r.t. the Gaussian state. That is to say, the calculability of the ansatz allows us to compute the expectation value of observables such as correlation functions in terms of a finite number of Gaussian expectation values. In particular, the terms proportional to $s^j$ in the non-Gaussian $n$-point correlation function correspond to $(n+m(j-1))$-point  Gaussian correlators, where $j=0,\ldots,n$ and $m$ is the power associated to the operator $\cB=\pi\phi^m$.

\section{Gaussian Variational Approach to the BH model}\label{sec:gva}
The Bose-Hubbard model  describes a system of interacting spinless bosons on the lattice. It is a theoretical description in a wide variety of contexts such as interacting ultracold atoms in optical lattices, ${}^{4}$He in various confined geometries and granular superconductors \cite{Bloch2008}. Its Hamiltonian in 1D is given by:
\begin{equation}
\label{BHH}
H_{\mathrm{BH}}=-t \sum_{<i, j>}\left(b_{i}^{\dagger} b_{j}+b_{i} b_{j}^{\dagger}\right) +\frac{U}{2} \sum_{t}\left(n_{i}\left(n_{i}-1\right)\right)-\mu \sum_{i} n_{i} \, .
\end{equation}
Here,  $< i, j >$ denotes nearest neighbours,  $[b_i, b_{j}^{\dagger}] = \delta_{ij}$  and ${n}_{i}={b}_{i}^{\dagger} {b}_{i}$ is the number operator. In the first term, the kinetic energy,   $t$ is the hopping amplitude. The second term stands for the on-site interaction with repulsion strength $U \;  (>0)$. For convenience, we include the chemical potential, $\mu$.

The dynamics of the BH model is given by the interplay between boson tunneling  (with amplitude $t$), and the repulsion between two bosons on the same site (of energy $U$). At fixed density, for small $U/t$,  the bosons can be considered as nearly free, so at low temperatures $T$ they condense into a superfluid with macroscopic occupation of the zero momentum single-particle state. In the opposite limit of large $U/t$, the repulsion between the bosons localizes them into a Mott insulator.  The Mott insulator is a state adiabatically connected to the product state with one boson in each potential minimum of the lattice. Interestingly, there is a second order quantum phase transition between these states which occurs at a critical value of $U/t$. This transition is driven by phase fluctuations and
belongs to the $XY$ universality class. In addition, there is another universality class for the Mott transition. A transition that occurs by changing the chemical potential (density), which is driven by density fluctuations and belongs to the mean-field universality class \cite{Fisher1989BosonLA,Cazalilla_2011}.

\medskip
Here we discuss the Gaussian varational ansatze (GVA) for the BH model in $D=1$.   For that, it is convenient to write the Hamiltonian in momentum space:
\be
{H}=\sum_{k} \varepsilon_{k} {b}_{k}^{\dagger} {b}_{k}+\frac{U}{2 N} \sum_{k, p, q} {b}_{k+q}^{\dagger} {b}_{p-q}^{\dagger} {b}_{k} {b}_{p}
\ ,
\ee
where  ${b}_{k}=\frac{1}{\sqrt{N}} \sum_{i} e^{-i k j} {b}_{j}$, with $[b_k,b^{\dagger}_{k'}]=\delta_{k,k'}$ 
and
\be
\varepsilon_{k}=  -2 t \cos \frac{2 \pi k}{N}-\mu,
\ee
refers to the (non-interacting) dispersion relation with $N$ the number of lattice sites. The generalization to $D>1$ is trivial.  
\\

The GVA is based on building a Gaussian variational trial state given by
\be
\label{GVA}
     | \Psi (\lambda;\beta_0)\ra_{G} = U_G | \Omega \rangle 
\ ,
\end{equation}
with $| \Omega\ra$ being the trivial vacuum $b_k | \Omega \rangle =0$ and
\be
    U_G = \cD(\beta_0) \; \cS(\lambda)
    \ .
\ee
Here $\cS$ a squeezed operator and $\cD$ a displacement operator (which accounts for boson condensation in the broken symmetry phase) given by
\begin{align}
    \cS(\lambda) &\equiv \exp\left(\frac{1}{2} \sum_k \lambda_k ( b_{-k}^\dagger b_k^\dagger - b_{-k} b_k ) \right) \, .\\ \nonumber
    \cD(\beta_0) &\equiv \exp\left(\beta_{0} ({b}_{0}^{\dagger}-{b}_{0})\right)\, . 
\end{align}

We note that $ \cS(\lambda)$ is particular case of Eq. \eqref{eq:generalized_cs} where $\Lambda_{k\, l}$ is diagonal in the mode basis with $\Lambda_{k\, l}\equiv \lambda_k\, \delta_{k+l,\, 0}$.
The unitary transformation in \eqref{GVA} yields a canonical linear transformation (Bogoliubov transformation) on the field operators given by
\be
U_G^{\dagger}\, b_k^\dagger\,  U_G = u_k\, b_k^\dagger - v_k\, b_{-k}+ \delta_{k0} \beta_0
\, ,
\ee
 where $u_k\equiv\cosh (\lambda_k)$ and $v_k \equiv \sinh (\lambda_k)$
The GVA amounts to finding the optimal variational parameters that minimize the ground state energy, that is
\be
\cE_G= \min_{ \beta_0, \lambda_k}   \left \{ \cE_G(\beta_0,\lambda)= \la \Omega | U_G^\dagger  \,  H  \, U_G | \Omega \ra \right \}\, ,
\ee
with
\begin{align}
    \cE_G(\beta_0,\lambda) &= \varepsilon_0 \beta_0^2  
+ \sum_k \varepsilon_k v_k^2\, \\ \nonumber
&+\frac{U}{2N} \Big [\big (\sum_k u_k v_k \big )^2 
+2 \big ( \sum_k v_k^2 \big)^2
 +4 \beta_0^2 \sum_k v_k^2\\ \nonumber
 &- 2 \beta_0^2 \sum_k u_k v_k +\beta_0^4\Big]\, ,
 \label{eq:energy_func}
\end{align}
upon which a numerical minimization must be carried out.  It is, however, instructive to consider some previous analysis. By setting $\lambda_k=0$, we obtain a coherent state ansantz ($\mathcal{S}(\lambda) = 1$).  In this case, the  energy  is minimized for $\left|\beta_{0}\right|$ equal to 
\begin{equation}
\beta_{0}^{\mathrm{c}}:=\sqrt{-\varepsilon_{0} N / U}
\, ,
\end{equation}
leading to the expectation value
\begin{equation}
\label{Egscoh}
E_{\left|\beta_{0}^{c}\right\rangle}=\left\langle\beta_{0}^{\mathrm{c}}|{H}| \beta_{0}^{\mathrm{c}}\right\rangle=\varepsilon_{0}\left|\beta_{0}^{\mathrm{c}}\right|^{2}+\frac{U}{2 N}\left|\beta_{0}^{\mathrm{c}}\right|^{4}=-\frac{\epsilon_{0}^{2} N}{2 U} \; .
\end{equation}

While the Gaussian ansatz \eqref{GVA} nicely captures some features of the BH ground state in the superfluid phase. the \eqref{GVA} wavefunction fails to capture the essential features of the Mott insulating phase. At zero order in $t/U$ in the Gutzwiller approximation \cite{PhysRevB.44.10328}, the Mott-phase wavefunction is given by
\be
|\Psi_{\rm Mott}\ra^{(0)}\equiv \prod_{k=1}^{N}\, \frac{1}{\sqrt{n_0!}}\, (b_k^{\dagger})^{n_0}|\Omega\ra\, ,
\ee
which it is obviously a non-Gaussian state ($n_0$ is the number of bosons on each site).

\section{NLCT-Wavefunctions for the the Bose-Hubbard model}\label{sec:nlct_bh}
Given the discussion above, here we apply the NLCT denoted by $\cB=\pi\, \phi^2$ to the BH model. Before entering into many details, we discuss on the suitability of this choice for the problem at hand.

The low energy physics in the vicinity of the critical point between the superfluid and Mott insulator driven by the ratio $t/U$ is described by a quantum field theory with an emergent
Lorentz invariant structure \cite{sachdev2011}. In this low energy effective theory,  the speed of ‘light’, $c$, is given by the speed of sound in the superfluid phase. In terms of the long-wavelength boson annihilation operator $\psi$, the Euclidean time action for the field theory is given by the quartic self-interacting scalar theory
\begin{align}
    S=\int d\tau dx \left(c^2 \vert \partial_x\psi\vert^2 + \vert \partial_{\tau}\psi\vert^2 + w\vert \psi\vert^2 + u\vert\psi\vert^4\right)\, ,
\end{align}
where $u$ is the coupling which tunes the system across the
quantum phase transition at some $u$ = $u_c$. For $u > u_c$, which corresponds to the Mott insulator,
the field theory has a mass gap and no symmetry is broken. The gapped particle and anti-particle states associated with the field operator $\psi$ correspond to the ‘particle’ and ‘hole’ excitations of the Mott insulator. These can be used as a starting point for a quasiparticle theory of the dynamics of the Mott insulator. The other phase with $u < u_c$ corresponds to the superfluid where the global U(1) symmetry of $S$ is broken and a quasiclassical Gaussian theory of the superfluid phase is possible.

This phase transition in the $\psi^4$ theory has been investigated through NLCT. More concretely, the $\cB=\pi\, \phi^2$ has first been used in \cite{Polley:1989wf, Ritschel:1990zs, IbanezMeier:1991hm},and lately in the context of continuous tensor networks \cite{Fernandez-Melgarejo:2019sjo, Fernandez-Melgarejo:2020fzw} and holography \cite{Fernandez-Melgarejo:2021mza}. It is thus sensible to apply this transformation to our problem, by remarking that the choice for different transformations must include a justification for the regimes of the theory one is interested to study. Furthermore, let us elaborate on the  effect of the NLCT transformation on wavefunctionals. In \cite{Ritschel:1990zs, Fernandez-Melgarejo:2020fzw} it is shown that $\cU$ generates a translation of the argument in the configuration space of the theory that symbolically reads  
\begin{equation}
   \Psi[\phi]_{NG}=\Psi_G[\phi-s\, \Phi]=\Psi_G[\phi-s\, \Phi]\, ,
   \label{eq:nlct_wavefunction}
\end{equation}
with $\Phi = h\, \phi^2$. In compact notation, writing $\Psi_G[\phi]=\exp\left[-\frac{1}{4} (\phi\,  \cdot\, {\bf G}^{-1}\, \cdot\,  \phi)\right]$, with ${\bf G}^{-1}$ the Gaussian kernel defining the correlation matrix of Gaussian states, the above identity can be cast as 
\begin{align}
    \Psi_{NG}[\phi]&=\exp\left[-\frac{1}{4} (\phi-s\, \Phi)\,  \cdot\,  {\bf G}^{-1}\, \cdot\,  (\phi-s\, \Phi)\right]\\ \nonumber
    &=\exp\Big[-\frac{1}{4} (\phi\,  \cdot\,  {\bf G}^{-1}\,  \cdot\,  \phi) -2\ s \, (\phi\,  \cdot\, {\bf G}^{-1}\,  \cdot\,  \Phi) \\ \nonumber
    &+ s^2\, (\Phi\, \cdot\,  {\bf G}^{-1}\,  \cdot\,  \Phi)\Big]\, 
\end{align}
from which one immediately infers that the new wavefunctional has been enhanced with variational \textit{skewness}  and \textit{kurtosis} terms (the term proportional to $s$ and $s^2$ respectively) that cannot be captured by a Gaussian ansatz. For esxample, in \cite{Damski_2015}, these terms have been characterized under different regimes of the Bose-Hubbard model. 

The bosonic field transformation \eqref{trasfields} is explicitly given by
\begin{align}
    \Phi_k
=&\
h_{k p_1p_2}\, \phi_{p_1}\phi_{p_2}\, \times  \delta_{k,p_1+p_2}', ,\\ \nonumber
\Pi_k
=&\
-2h_{p_1 k p_2}\, \pi_{p_1}\phi_{p_2}\, \times  \delta_{k,p_1+p_2}\, ,
\end{align}
where a summation index convention is assumed. In order to proceed it is convenient to provide an ansatz for the variational parameters $h_{p,q_1,q_2}$ that fulfills the truncation constraints in \eqref{eq:constraint}. This can be easily achieved by taking  the decomposition 
\begin{align}
h_{p,q_1,q_2} = \eta(p)\cdot \zeta(q_1) \cdot  \zeta(q_2)\, ,
\end{align}
where it is imposed that $\eta(p)\cdot \zeta(p) = 0$, \emph{i.e.}, the domains of momenta, where $\eta$ and $\zeta$ are different from zero have to be disjoint, up to sets of measure
zero. A suitable ansatz for $\eta$ and $\zeta$ is given by 
\begin{eqnarray}
\label{eq:cutoff_ansatz}
\eta(p)&=&\Gamma((p/\Delta_1)^2)\, , \\ \nonumber
\zeta(q_i) &=& \left[\Gamma((\Delta_1/q_i)^2)-\Gamma((\Delta_2/q_i)^2)\right]\, ,
\end{eqnarray}
where  $\Delta_{1}$ and $\Delta_2$, are variationally optimized, coupling dependent momentum cutoffs and $\Gamma(x) = \Theta(1-|x|)$, with $\Theta$ the Heaviside step function.

With this, our objective is to compute 
\be
\cE_{NG}= \min_{ \beta_0, \lambda, h}   \left \{ \cE_{NG}(\beta_0,\lambda, h)= \la \Psi | \cU^\dagger  \,  H  \, \cU | \Psi \ra_{NG} \right \}\, .
\ee
To facilitate the task we describe, below, the calculation of the energy with the non-Gaussian ansatz on a term-by-term basis.

\subsection{Kinetic Term.}
For the \emph{kinetic} term  we obtain
\begin{align}
\cE^{\rm kin}_{NG} &= \sum_{k}\varepsilon_k\left( \expval{b^\dagger_{k}b_k }
+s^2\expval{B^\dagger_{k}B_{k}}\right) 
= \cE^{\rm kin}_{G}  +s^2\sum_{k}\varepsilon_k\expval{B^\dagger_{k}B_{k}}
\, ,
\end{align}
where $\cE^{\rm kin}_{G} =\varepsilon_0 \beta_0^2  
+ \sum_k \varepsilon_k v_k^2$ and
\begin{align}
B_k \equiv \frac{1}{\sqrt2}(\Phi_k+i\Pi_k)
\ ,
\quad
B^\dagger_k \equiv \frac{1}{\sqrt2}(\Phi_{-k}-i\Pi_{-k})
\ .
\end{align}

After a lengthy albeit straightforward calculation the result for the kinetic term can be written as
\begin{align}
\cE^{\rm kin}_{NG} &= \cE^{\rm kin}_{G} + s^2(2 \chi_2 + \varepsilon_0\chi_1^2)\\ \nonumber
&=\varepsilon_0 (\beta_0^2  +s^2\, \chi_1^2)
+ \sum_k \varepsilon_k v_k^2 + 2\, s^2 \chi_2 
\, ,
\end{align}
where, defining $G_k=\frac{1}{2}\, \exp\left(-2\lambda_k\right)$, $\chi_i$ are given by
\begin{align}\label{eq:kinetic_chis}
    \chi_1 &= \sum_{p}  h_{0,p,p}\, G_p\, , \\ 
    \chi_2 &= \sum_{p,q}\, \varepsilon_{p+q}\left(  h^2_{p+q,p,q}\, G_p\, G_q\,
	+\frac{1}{8}  h_{p,p+q,q}^2\, \frac{G_q}{G_p}
	\right)
\, .
\end{align}

\subsection{Interaction Term.}
For the interaction term, we obtain
\begin{align}
\expval{ b^\dagger_{k+r}b^\dagger_{l-r}b_k b_l}_{NG} &=
\expval{b^\dagger_{k+r}b^\dagger_{l-r}b_k b_l}\\ \nonumber
&+s^2\Big[
	\expval{B^\dagger_{k+r}B^\dagger_{l-r}b_k b_l}
	+\expval{B^\dagger_{k+r}b^\dagger_{l-r}B_k b_l}\\ \nonumber
	&+\expval{B^\dagger_{k+r}b^\dagger_{l-r}b_k B_l}
	+\expval{b^\dagger_{k+r}B^\dagger_{l-r}B_k b_l}\\ \nonumber
	&+\expval{b^\dagger_{k+r}B^\dagger_{l-r}b_k B_l}
	+\expval{b^\dagger_{k+r}b^\dagger_{l-r}B_k B_l}\Big]\\ \nonumber
&+s^4 \expval{B^\dagger_{k+r}B^\dagger_{l-r}B_k B_l}
\, .
\end{align}

As it will be justified below, our calculations will be carried out in such a way that $s^2 \gg s^4$ notwithstanding that a large value of the interaction strength $U$ can be taken. Due to this, the $\cE^{\rm int}_{NG}$ will take the form
\begin{align}
    \cE^{\rm int}_{NG}= \cE^{\rm int}_{G} + s^2\, \frac{U}{2N}\, \Sigma_{\rm int}\, ,
\end{align}
where 
\begin{align}
    \cE^{\rm int}_{G} &= \frac{U}{2N} \Big [\big (\sum_k u_k v_k \big )^2 +2 \big ( \sum_k v_k^2 \big)^2
 +4 \beta_0^2 \sum_k v_k^2\\ \nonumber
 &- 2 \beta_0^2 \sum_k u_k v_k +\beta_0^4\Big]
 \, ,
\end{align}
and
\begin{align}
    \Sigma_{\rm int}=\Sigma_{0} + \chi_1\, \Sigma_{1} + \chi_1^2\Sigma_{2}\, . 
\end{align}
The explicit expressions for the $\Sigma$'s in terms of the variational parameters  are rather lengthy so we refer the reader to Appendix to find them.

\subsection{Final expression}

At this point, and following \cite{Ritschel:1990zs} it is convenient to write the non-Gaussian energy density in terms of a different set of variables. To this end, we note that after the transformation $\pi\, \phi^2$,  $\expval{b_k}_{NG} = \beta_0+s\, \chi_1$ depends on several parameters of the ansatz. We define
a new parameter $\psi_c\equiv s\chi_1$, and, thus, one of the original variables can be eliminated. The resulting energy density is in general different from the Gaussian case and can be written as
\begin{align}\label{eq:energy_functional_psi}
    \cE_{NG} = \cE_{G} + 2s^2\chi_2 + \psi_c^2\epsilon_0 + \frac{U}{2N} \Big [s^2\Sigma_0 + s\psi_c\Sigma_1 + \psi_c^2\Sigma_2\Big]
 \, .
\end{align}
From this last expression it is straightforward to find an optimal $s$  for a fixed value of $\psi_c$ and optimized values of $\Delta_1$ and $\Delta_2$ by,
\begin{eqnarray}\label{eq:optimal_s}
\begin{aligned}
     \frac{\partial }{\partial s}\,  \cE_{NG}\Bigr|_{\psi_c} &= 4s\chi_2 + 2\frac{U}{2N}s\Sigma_0 + \frac{U}{2N}\psi_c\Sigma_1=0\, ,\\ 
     \bar{s} & = -\frac{(U/2N)\Sigma_1}{(2\chi_2 + (U/2N)\Sigma_0)}\, \psi_c\, .
     \end{aligned}
    \end{eqnarray}
Analogously, it is possible to find a set of equations for finding the optimal values of the variational parameters $\Delta_1$ and $\Delta_2$ by
\begin{align}
    \frac{\partial }{\partial h_{k+q,k,q}}\,  \cE_{NG}\Bigr|_{\psi_c} = 0
 \, .
 \end{align}
 In the case of the $\lambda\phi^4$ theory these are a kind of Fredholm integral equations \cite{IbanezMeier:1991hm} that can be solved numerically. In our case the equations are rather involved so we use an alternative numerical procedure to find the optimal values. This will be described in the next section.

\section{Results}
\label{sec:results}

We carry out the optimization of the ground state energy functional in \eqref{eq:energy_functional_psi} following the lines exposed in \cite{Polley:1989wf, Ritschel:1990zs, Fernandez-Melgarejo:2020fzw}. This consists in obtaining the variational parameters of the ansatz in two consecutive steps. First, we optimize the Gaussian ansatz to obtain the optimal $\bar{\beta}_0,\, \bar{\lambda}$ parameters. Once these parameters are obtained, they are fixed in order to carry out the optimization of the non-Gaussian parameters in a separate fashion, that is, 
\begin{equation}
    \cE_{NG}(\lambda,h) \quad \to \quad \cE_{NG}(\bar{\lambda}; h)=\cE_G(\bar{\lambda}) + \delta \cE(\bar{\lambda}; h)\, ,
\end{equation}
where $\delta \cE(\bar{\lambda}; h)\equiv \delta \cE(\bar{\lambda}; \chi, \Sigma)\equiv \delta \cE(\bar{\lambda}; \Delta_1, \, \Delta_2)$. To further simplify the process, we fix 
 $\Delta_2$ to its maximum allowed value of $2 \pi (N-1)/N$ leaving the parameter $\Delta_1$ (which can take values from $0$ to $2 \pi (N-1)/N$) as the only one parameter needed to determine the ground state energy.
 
After carrying out the optimization procedure described above, we note that  the $\chi$'s and $\Sigma$'s in \eqref{eq:energy_functional_psi},  directly depend on the total number $M$ of non-linearly modified modes $\phi_k$. Intuitively, this means that $h_{k,p,q}$ weighs how much a low energy mode  $\phi_k$ (with $k \leq \Delta_1$)  is modified by high energy modes $\phi_p$ and $\phi_q$  (with $\Delta_1\leq p,\,q \leq \Delta_2$). According to our ansatz, a given value, for instance, the optimal value of the variational parameter $\Delta_1$, defines the total amount of the non-linearly modified modes as $M=(\Delta_1/2\pi)\, N < N$. 
 
 Following \cite{Ritschel:1989ib}, such dependence can be written as 
 \begin{eqnarray}
             \chi_2 = a_2\, M^{\, \alpha_2}\, ,\quad \Sigma_0 = c_0\,  M^{\, \gamma_0}\, ,\\ \nonumber
             \Sigma_1= c_1\, M^{\, \gamma_1}\, ,\quad \Sigma_2 = c_2\, M^{\, \gamma_2}\, ,
 \end{eqnarray}
 where $a_2\, ,c_0\, , c_1$ and $c_2$ are slowly varying functions of $t, U$ and $\mu$. Given the definition of the independent parameter $\psi_c$, that is, a non Gaussian correction to the Gaussian condensate, in terms of $\chi_1$, it is sensible to assume that $\psi_c = b\, M^{1/2}$. That is to say, assuming that $b$ is an $\mathcal{O}(1)$ constant, the non-Gaussian correction to the condensate value is proportional to $M^{1/2}$ in such a way that its correction to the energy density is $\propto \psi_c^2 \sim M$. Consequently, we obtain
\begin{widetext}
\begin{align}
    \cE_{NG}= \varepsilon_{G} N + 2s^2\, a_2\, M^{\alpha_2} + b^2\, M \epsilon_0 + \frac{U}{2N} \Big [c_0\, s^2\, M^{\, \gamma_0} + b\, c_1\, s\,  M^{\, \gamma_1 +1/2} + b^2\, c_2\, M^{\, \gamma_2 +1}\Big]
 \, ,
\end{align}
\end{widetext}
where $\varepsilon_{G}$ is the energy density of the Gaussian ansatz. In order to get further insights, we use that, after optimizing the energy functional \eqref{eq:energy_functional_psi}, a numerical analysis yields  $\alpha_2=\gamma_1=0$, $\gamma_0 = \gamma_2 =1$, $a_2, c_1, c_2 < 0$ and $c_0>0$ which leaves
\begin{widetext}
\begin{align}
    \cE_{NG}&= \varepsilon_{G} N + 2s^2\, a_2 + b^2\, M \epsilon_0 + \frac{U}{2N} \Big [c_0\, s^2\, M + b\, c_1\, s\,  M^{\, 1/2} + b^2\, c_2\, M^{2}\Big]
    \, ,
    \\ 
 \bar{s}& = -\frac{U\, b\, c_1}{(8 N\, a_2 + 2 U\, c_0\, M)}\, M^{1/2}\, .
\end{align}
\end{widetext}

Thus, defining $z=\Delta_1/2\pi$, in the limit where $z \ll 1$. \emph{i.e.}, for  $M\ll N$ ,the optimal $s$ can be written as
\begin{equation}
     \bar{s} = -\frac{U}{2 N}\, \left(\frac{b\, c_1}{4 a_2}\right)\, \sqrt{M} = -\frac{\alpha}{N}\, \sqrt{M}.
\end{equation}
This result implies that our truncated estimate of the ground state energy at order $s^{2}$ is justified and can be written as
\begin{widetext}
\begin{align}
    \frac{\cE_{NG}}{N}&= \varepsilon_{G}  + 2\alpha^2\, a_2\, \left(\frac{M}{N^3}\right) + b^2\, \left(\frac{M}{N}\right)\epsilon_0 + \frac{U}{2} \Big [c_0\, \alpha^2\, \left(\frac{M}{N^2}\right)^2 - b\, c_1\, \alpha\,  \left(\frac{M}{N^2}\right) + b^2\, c_2\, \left(\frac{M}{N}\right)^2 \Big]
 \, .
\end{align}
\end{widetext}

More explicitly, for $\mu >0$, the leading contributions to the variation of the energy density estimation w.r.t. the Gaussian case reads 

\begin{align}
   \frac{\delta\, \cE}{N}&=   b^2\, z\, \Big [\epsilon_0 + \frac{U}{2}  c_2\, z \Big] < 0\, ,\\ 
    \frac{\delta\, \cE}{N} &=   -b^2\, z\, U\, \Big [\left(2 \left(\frac{t}{U}\right) + \left(\frac{\mu}{U}\right)\right) + \frac{|c_2|}{2}  \, z \Big] 
 \, .
 \label{eq:qualit_ansatz}
\end{align}
 
 The last expression allows us to make some specific quantitative predictions for the behaviour of the ansatz at different points of the phase diagram. For instance, for a fixed $t/U \ll 1$, one might expect that
 \begin{align}
     \frac{\delta\, \cE}{N}  \sim   -b^2\, z\,  \Big [ \mu + U\, \frac{|c_2|}{2}  \, z \Big] 
 \, .
 \label{eq:qualit_ansatz_f1}
\end{align}

\begin{figure}[t]
\centering 
\includegraphics[width=0.5\textwidth]{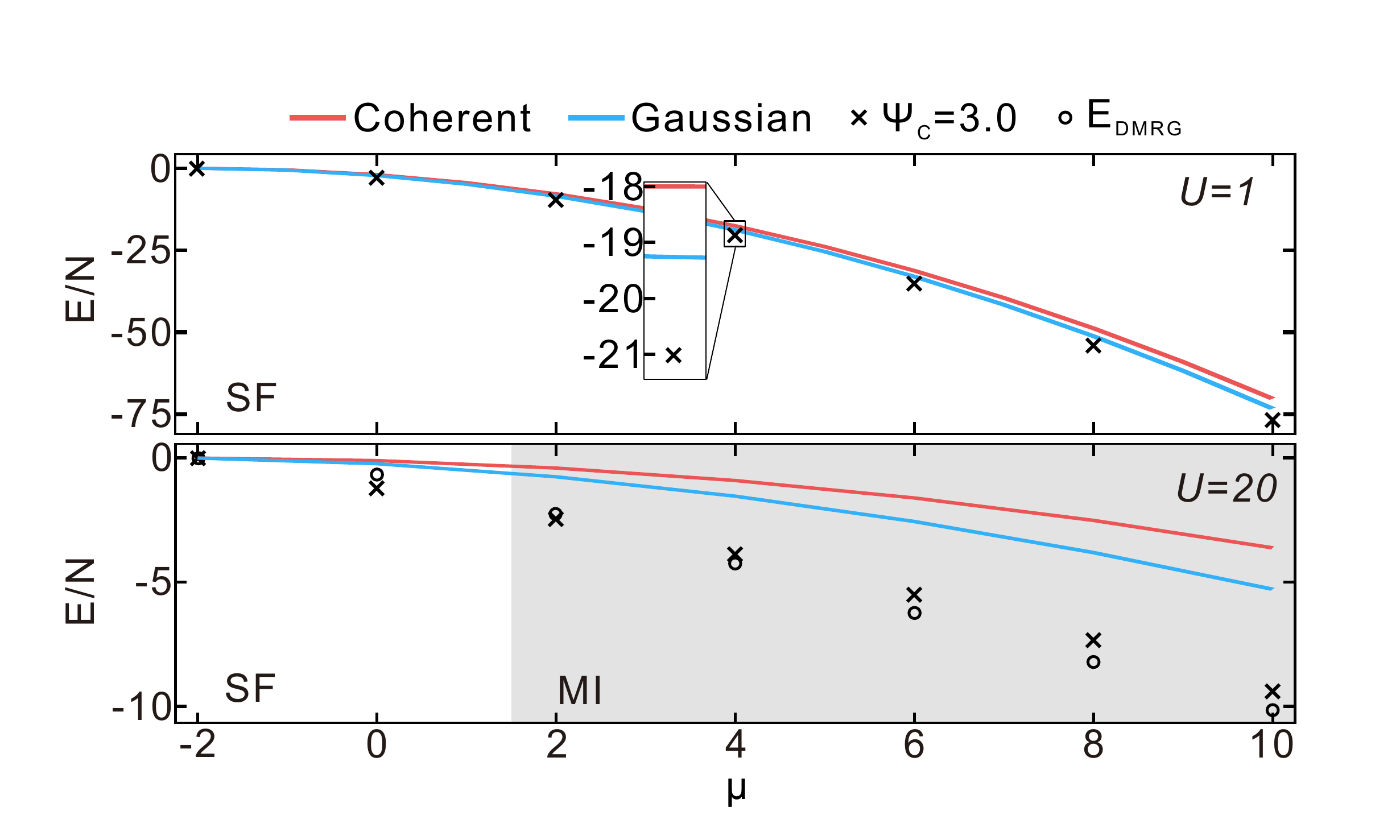} 
\caption{\textit{Ground state energies with respect to $\mu$.} For $N=32$ and $t=1$, we compare the performance of our approach with a Coherent state and the Gaussian state for weak and strong interaction strengths $U=1$ and $U=20$ respectively.  DMRG results were taken from \cite{Guaita_2019} where authors  computed  the energy density for finite systems with open
boundary conditions and then extrapolated to the thermodynamic limit. } 
\label{fig:GE_mu} 
\end{figure}

With Eq. \eqref{eq:energy_functional_psi},  we compute the predictions of our ansatz for the  ground state energy density. The performance of our method is tested by comparing this quantity with the ones obtained through a Coherent state, a Gaussian state and DMRG \cite{Guaita_2019} (see Fig. \ref{fig:GE_mu}). Both the energy densities obtained through a coherent and a Gaussian state are higher than the DMRG one, especially as one deeps into the MI phase. While, the Gaussian variational family provides a consistent class to approximate the ground state of the Bose-Hubbard model in the superfluid phase, our non-Gaussian approach is capable of both, \textit{i)} providing a consistent non-perturbative ($U$-independent) truncation of the bosonic opèrator growth under a non-Gaussian transformation and \textit{ii)}  predicting a much better approximation to the ground state energy  than previous methods. This is quite manifest in the MI phase, where the energy estimation improvement over the Gaussian ansatz is qualitatively explained by \eqref{eq:qualit_ansatz_f1}. Remarkably, the non-Gaussian ansatz performance on the superfluid phase equals the one obtained by the Gaussian ansatz, as far as  the fraction of non-linearly modified modes $z\sim 0$ in this phase.

\begin{figure}[t]
\centering 
\includegraphics[width=0.5\textwidth]{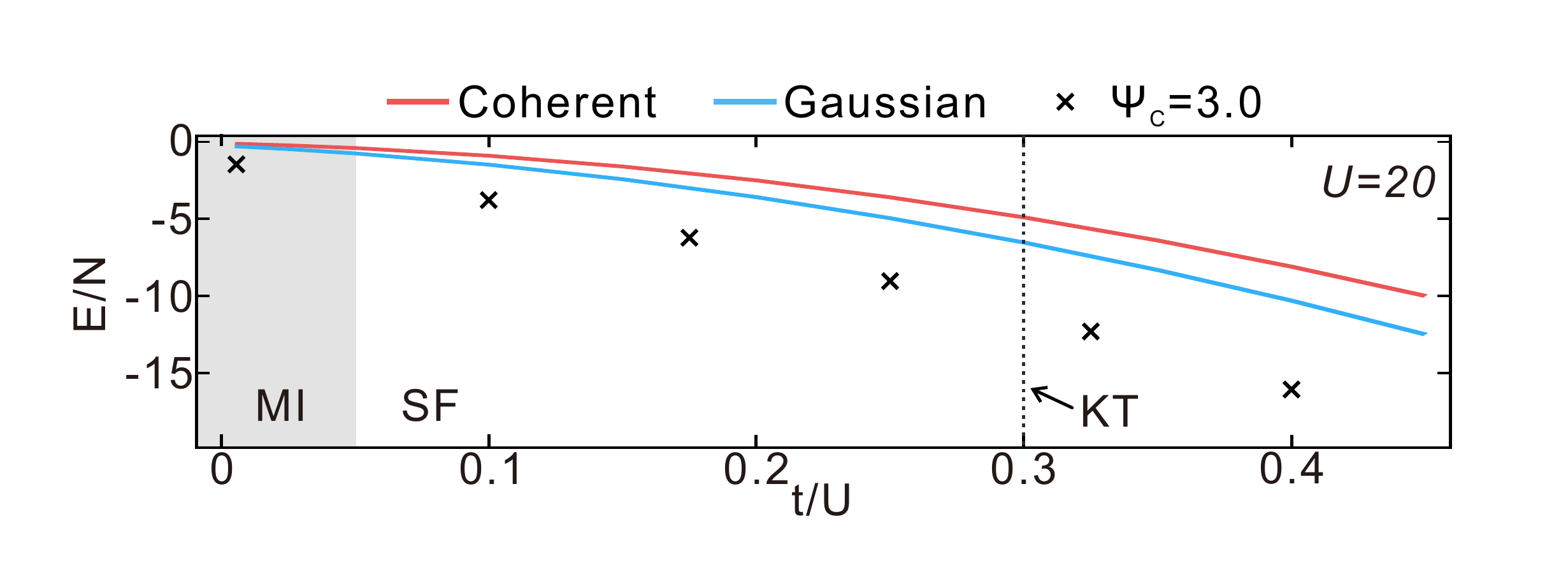} 
\caption{\textit{Ground state energy with respect to $t/U$.} For $N=32,U=20,\mu=2$, we compare the energy density predictions for the Bose-Hubbard model for fixed $\mu$ as a function of $t/U$. } 
\label{fig:GE_t} 
\end{figure}

We also show in Fig. \ref{fig:GE_t} the energy density computed through the NLCT-ansatz, for varying $t/U$ with $\mu$ fixed. The numerical results show agreement with the qualitative predictions yielded by \eqref{eq:qualit_ansatz} and with results in \cite{Damski_2015} where authors employed a high-order perturbative expansion to characterize the ground state of the Mott phase of the 1D Bose–Hubbard model. 

Results in Fig \eqref{fig:GE_mu} and Fig \eqref{fig:GE_t} are shown for a total lattice sites $N=32$. We have numerically checked that the results are  similar for larger values of $N\, (N=128,\, N=512)$. This might be expected in terms of Eq. \eqref{eq:qualit_ansatz}. There, the improvement over the Gaussian prediction is given in terms of the ratio $z=M/N$. This is uniquely determined by a variational parameter $\Delta_1$ which only depends on the point of the phase diagram under consideration. In our simulations, in the Mott phase, $\Delta_1$ has been established to have an average value of $2\pi/5$ which implies that $M\sim N/5$.  This fixes $\psi_c = b\, \sqrt{N/5}|_{N=32}\sim 3.0$, $\psi_c = b\, \sqrt{N/5}|_{N=128}\sim 6.0$  and $\psi_c = b\, \sqrt{N/5}|_{N=512}\sim 12.0$. Here, we have used that $b \sim \mathcal{O}(1)$ parameter.
Thus, numerical results based on a construction of the wavefunction  ansatz that is fully non-perturbative, strongly suggests that our method accurately approximates ground state energy per lattice site across the critical point.

\section{Discussion and Outlook}\label{sec:discussion}
In this work, we have used a class of variational non-Gaussian wavefunctions which extend Gaussian states by means of NLCT. This technique was proposed as a variational non-perturbative method to study phase transitions in QFT \cite{Polley:1989wf,Ritschel:1989ib,Ritschel:1990zs,IbanezMeier:1991hm} and has been recently developed for building controlled settings to study the AdS/CFT holographic duality through tensor networks \cite{Fernandez-Melgarejo:2019sjo,Fernandez-Melgarejo:2020fzw,Fernandez-Melgarejo:2020utg,Fernandez-Melgarejo:2021mza}.  Here, the NLCT-wavefunction  method has been applied to the 1D-Bose-Hubbard model. We obtain a family of approximate ground states for arbitrarily large values of the interaction strength. Our results show that a particular class of these states is able to sensibly improve the ground state energy estimation when the system is in the Mott phase \cite{Guaita_2019}.
 
We have used the ground state energy as a test-bed to benchmark the validity of the NLCT approach. It has been left as a future problem to address, to what extent the full non-perturbative structure of the ground state of the BH model is being captured by our concrete ansatz. This connects with the question of choosing the most appropiate non-Gaussian operator $\cB = \pi \phi^{n}$. The study of connected correlation functions, which can be systematically addressed through Eq. \eqref{eq:corr_funcs} (see \cite{Fernandez-Melgarejo:2020fzw,Fernandez-Melgarejo:2021mza}) might shed light on this. This amounts to quantify the \emph{skewness} and \emph{kurtosis} \cite{Damski_2015} of the system in different regimes and compare them with our ansatz. An additional benchmark to test the NLCT-ansatz  is the prediction of the critical point $(t/U)|_{\rm crit}$ for the superfluid-Mott insulator transition. For this, it is necessary to perform the explicit evaluation of the $s^{4}$-term in the ground state energy functional, Eq. \eqref{eq:energy_functional_psi}. This procedure has been used  in the $\lambda\, \phi^{4}$ theory \cite{Ritschel:1990zs} (see reference \cite{Milsted:2013rxa} for a comparison with other methods including tensor networks). For the BH model in particular, a comparison with \cite{Ejima_2012} and \cite{Lindinger_2019}) might further help to elucidate to which extent our ansatz covers the non-perturbative structure of the BH ground state.

 Another interesting possibility offered by our method is given by the translational effect on the wavefunctional argument exhibited in \eqref{eq:nlct_wavefunction} and which has been partially explored in \cite{Chen:2020ild,Fernandez-Melgarejo:2020utg, Fernandez-Melgarejo:2021mza}. Given the calculability
bonanza shown by the variational ansatze used in this paper, it is possible to compute the entanglement entropy of arbitrary regions in the BH ground state by using the prescription for the ground state of free theories. In free theories, the entanglement entropy is fully determined by the two-point correlation functions. For the interacting case, it is shown that  non-Gaussian contributions to the entanglement entropy can be obtained through a Gaussian prescription by replacing the Gaussian two point functions by their non-Gaussian counterparts obtained by the NLCT method \cite{Fernandez-Melgarejo:2020utg}.
 
 Finally, it is worth to mention possible extensions of the NLCT formalism to study out-of-equilibrium interacting bosons.  Specifically, it is interesting to understand how those systems respond to quenches. Extending the NLCT-procedure would allow us to study the spreading of correlations and entanglement in the strongly interacting regime of the Bose-Hubbard model \cite{L_uchli_2008,Bohrdt:2016vhv}. As the GVA possesses a well defined time-dependent extension, we expect that the structure of wavefunctions built from NLCT could help to establish a well defined extension of the non-Gaussian ansatze treated here in order to address these problems in the future.
 
\section*{Acknowledgements}
TQ would like to thank the financial support from China Scholarship Council and the guidance on the computations from Dr Junjie Zeng. He also thanks the hospitality at Instituto de Nanociencia y Materiales de Aragon during the initial stages of the project. JJFM and JMV thanks the financial support of Spanish Ministerio de Ciencia e Innovaci\'on PID2021-125700NA-C22.
DZ acknowledges the financial support of Spanish Ministerio de Ciencia e Innovaci\'on PID2020-115221GB-C41/AEI/10.13039/501100011033, the Gobierno de Araǵ\'on
(Grant E09-17R Q-MAD) and the CSIC Quantum Technologies Platform PTI-001. 
\medskip 

\appendix*
\section{Explicit form of $\Sigma $ terms}
For notational convenience  we denote $\tilde u_k = v^2_k + u^2_k$ and the index summation $\sum_{k,p,q}$ is assumed. With this, after a cumbersome albeit straightforward calculation, we have,
\begin{widetext}
\begin{eqnarray}
    \begin{aligned}
           2\, \Sigma_{0} &= 
           \Gamma_{p q}  \Big( 2 (\tilde{u}_k-1) - u_k v_k-\tilde{v}_k \Big)  + \Gamma_{p}^{q} \Big(\tilde{v}_k  + (\tilde{u}_k-1)\Big)  +\Gamma^{p}_{(k q)} \Big( 7 v_k v_q + ( u_k v_q + u_q v_k)\Big) - \Gamma_p^{(k q)} u_q v_k\\
            &+\Gamma^{p (k)}_{(q)} \Big(9 v_k v_q + ( u_k v_q-u_q v_k) + u_k u_q \Big)  
            + \Gamma^{p (q)}_{(k)}  \Big(3 v_k v_q  - u_k v_q + u_k u_q + v_k u_q\Big)  + \Gamma^{p (k q)}  \Big(13 v_k v_q + u_k u_q - u_q v_k\Big)
    \end{aligned}
\end{eqnarray}
\begin{eqnarray}
    \begin{aligned}
           \Sigma_{1} &= 
           -\Upsilon^{p (k)}_{(q)}  u_k v_q
            + 3 \Upsilon^{p (q)}_{(k)}   v_k v_q   + \Upsilon^{p (k q)}  \Big(10 v_k v_q + v_q v_k -8 v_k u_q - 6 u_k v_q\Big)
    \end{aligned}
\end{eqnarray}

\begin{eqnarray}
	\begin{aligned}
		4\, \Sigma_2=&  \Big(4(\tilde{u}_k -1) -4 u_k v_k \Big)
	\end{aligned}
\end{eqnarray}
\end{widetext}

where we have introduced the shorthand notation
\begin{eqnarray}
	\begin{aligned}
			\Upsilon^{r (l)}_{(m)}&= h_{l,l + m,m} \mathbb{C}^{r (l)}_{(m)}\, , \quad \Upsilon^{r (l m)}=h_{l+m,l,m} \mathbb{C}^{r (l m)}\, ,
			\end{aligned}
\end{eqnarray}

\begin{align}
			\Gamma_{l m} &= \mathbb{A}_{l m} h_{l+m,l,m}^2 \, ,\quad \Gamma_{l}^{m}= \mathbb{A}_{l}^{m}h_{l,l+m,m}^2\, \\ \nonumber
			\Gamma_{r}^{(l m)} &= h_{r,l+m,l} h_{r,r+m,m}\mathbb{C}_{r}^{(l m)}\, , \quad \Gamma^{r}_{(l m)}= h_{l+r,r,l} h_{m,r+m,r}\mathbb{C}^{r}_{(l m)}\, \\ 
			\Gamma^{r (l)}_{(m)} &= h_{k+l,r,l} h_{m,r+m,r}\mathbb{C}^{r (l)}_{(m)}\, ,\quad \Gamma^{r (l m)}=h_{l+r,r,l} h_{r+m,r,m}\mathbb{C}^{r (l m)} \nonumber
\end{align}

and

\begin{align}
			\mathbb{A}_{l m} &= G_l G_m\, ,\quad \mathbb{A}_{l}^{m}=\frac{G_l}{ G_m} \\ \nonumber
			\mathbb{C}_{r}^{(l m)}&= \frac{\sqrt{G_l G_m}}{G_r}\, , \quad
			\mathbb{C}^{r}_{(l m)}=\frac{G_r}{\sqrt{G_l G_m}}\\
			\mathbb{C}^{r(l)}_{(m)}&=G_r\frac{\sqrt{G_l}}{\sqrt{G_m}}\, \quad
			\mathbb{C}^{r (l m)}=G_r\sqrt{G_l G_m} \nonumber
\end{align}

\medskip

\bibliography{bib_bh}

\end{document}